\documentclass[aps,prb,twocolumn,amsmath,nofootinbib]{revtex4-1}

\usepackage{epsfig}
\usepackage[colorlinks,linkcolor=blue,anchorcolor=blue,
            citecolor=blue,urlcolor=blue,
            breaklinks=true]{hyperref}
\usepackage{graphics}
\usepackage{color}
\usepackage{dcolumn}% Align table columns on decimal %point
\usepackage{bm}% bold math
\usepackage{cases}
\usepackage{appendix}
\usepackage{lineno,hyperref}
\usepackage{ulem}

\begin{document}
\author{Run Cheng$^{1,2}$}
\author{Yong-Long Wang$^{2,3}$}
\email{wangyonglong@lyu.edu.cn}
\author{Hua Jiang$^{2}$}
\author{Xiao-Jun Liu$^{1}$}
\email{liuxiaojun@nju.edu.cn}
\author{Hong-Shi Zong$^{1,3,4}$}
\email{zonghs@nju.edu.cn}
\address{$^{1}$ Key Laboratory of Modern Acoustics, Department of Physics, Nanjing University, Nanjing 210093, P. R. China}
\address{$^{2}$ School of Physics and Electronic Engineering, Linyi University, Linyi 276005, P. R. China}
\address{$^{3}$ Joint Center for Particle, Nuclear Physics and Cosmology, Nanjing 210093, P. R. China}
\address{$^{4}$ State Key Laboratory of Theoretical Physics, Institute of Theoretical Physics, CAS, Beijing 100190, P. R. China}

\title{Geometric effects of a quarter of corrugated torus}
\begin{abstract}
In the spirit of the thin-layer quantization scheme, we give the effective Shr\"{o}dinger equation for a particle confined to a corrugated torus, in which the geometric potential is substantially changed by corrugation. We find the attractive wells reconstructed by the corrugation not being at identical depths, which is strikingly different from that of a corrugated nanotube, especially in the inner side of the torus. By numerically calculating the transmission probability, we find that the resonant tunneling peaks and the transmission gaps are merged and broadened by the corrugation of the inner side of torus. These results show that the quarter corrugated torus can be used not only to connect two tubes with different radiuses in different directions, but also to filter the particles with particular incident~energies.
\bigskip

% Electronic transport phenomena in thin film, 73.50.-h;
% Electron states at surfaces and interfaces, 73.20.-r;
% Quantum mechanics, 03.65.-w;
% Geometry, differential geometry, and topology, 02.40.-k;
\end{abstract}
\maketitle

\section{Introduction}
With the rapid development of nanotechnology, a variety of nanostructures with complex geometries are successfully fabricated, and  the related geometric effects {have attracted study by} both theoretical and experimental researchers \cite{Liu2018Emerging, Hamed2018Effect}. Specifically, the geometric effects are shown by the geometric potential~\cite{Costa1981Quantum, Jaffe2003Quantum}, geometric momentum~\cite{Liu2007Constraint, Liu2011Geometric}, geometric orbital angular momentum~\cite{Wang2017Geometric}, geometric gauge potential~\cite{Jaffe2003Quantum, Wang2018Geometric, Wang2018Erratum}, geometric magnetic moment~\cite{Brandt2015Induced,Wang2018Geometric} and so on. One of the most important geometric effects is the geometric potential that has been investigated widely. Researchers found that the geometric potential can be used to change the band-structure of the geometrically deformed systems~\cite{Aoki2001Electronic, Fujita2005Band, Aoki2005Electronic}, to generate localized surface states~\cite{Goldstone1992Bound, Cantele2000Topological, Encinosa2003Curvature, Taira2007Electronic, Taira2007Curvature, Ortix2010Effect} and prompt energy shifts~\cite{Encinosa1998Energy, Shima2009Geometry} for an electron confined to a curved surface;  {additionally, transport properties on a curved surface or space curve can be affected}~\cite{Marchi2005Coherent, Zhang2007Quantum, Cuoghi2009Surface, Shima2009Geometry}. Experimentally, the geometric potential~\cite{Szameit2010Geometric} and the geometric momentum~\cite{Schmidt2015Curvature} {have been} observed. These results provide evidence on the validity of the confining potential method.

A nanotube has nonvanishing curvature that contributes an attractive scalar potential for a~particle moving on the tube. When the nanotube is deformed~\cite{Goldstone1992Bound, Jaffe1999Dirac, Novakovic2011Transport, Moraes2016Geometric}, the geometric potential will be substantially reconstructed by the deformation. In other words, the deformation plays an important role in the effective quantum dynamics, like Dirac comb~\cite{Sakurai1993Modern}, with attractive wells. Particularly, the geometric potential is mainly induced by the deformation, such as the case for corrugations on the surface~\cite{Li2008Strain, Turner2010Vortices, Mutilin2014Microtubes, Wang2016Transmission}. The~corrugation extremely affects the geometric potential, and further influences the corresponding electronic transport. Usually, corrugations appear in the bend of a tube. To connect smoothly two tubes with different radiuses in different directions, we can consider a part torus with corrugation, a quarter corrugated torus particularly in the present paper.

This paper is organized as follows. In Section~\ref{Section2}, we deduce the effective quantum equation for a~particle confined to a periodically corrugated torus by the confining potential approach, and analyze the associated geometric potential. In Section~\ref{Section3}, we investigate the effects of the corrugations on transmission probability. Finally, in Section~\ref{Section4}, the conclusions are given.

\section{Quantum Dynamics of a Particle Confined to a Periodically Corrugated Torus}\label{Section2}
\begin{figure}[htbp]
\centering
\includegraphics[width=0.3\textwidth]{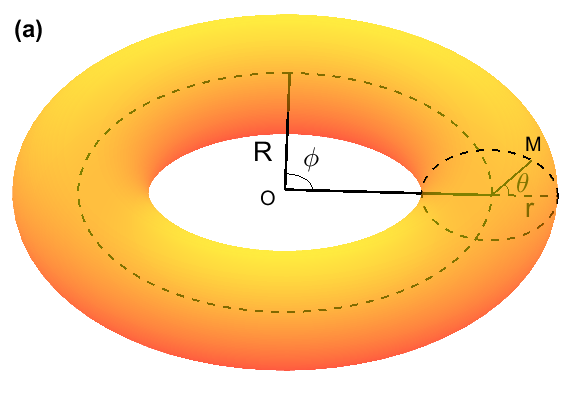}\\
\includegraphics[width=0.3\textwidth]{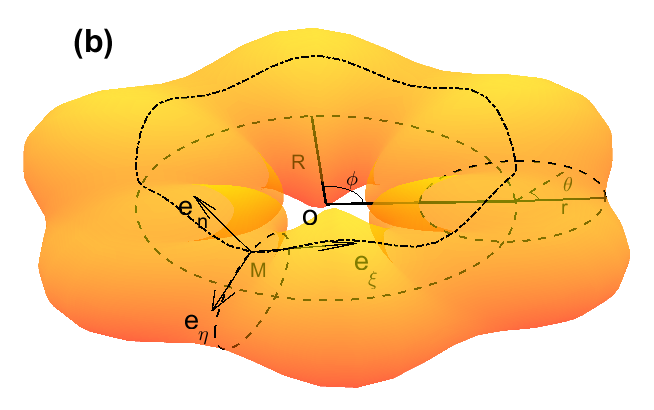}
\caption{(\textbf{a}) schematic of a torus with ${r}_0=(r_{x_0}, r_{y_0}, r_{z_0})$; (\textbf{b}) schematic of a corrugated torus described by ${r}=(r_x, r_y, r_z)$ with $n=6$. Here, $R$ and $\phi$ are the large radius and azimuthal angle, and $r$ and $\theta$ are the small radius and polar angle.}\label{Surf1}
\end{figure}

A torus reconstructed by corrugation (see Figure~\ref{Surf1}b) that is parametrized by
\begin{equation}\label{Surf}
{r}=(r_x, r_y, r_z),
\end{equation}
where $r_x, r_y$ and $r_z$ are
\begin{equation}
\begin{split}
& r_x=r_{x_0}+r_x^{\prime},\\
& r_y=r_{y_0}+r_y^{\prime},\\
& r_z=r_{z_0}+r_z^{\prime},
\end{split}
\end{equation}
respectively, with
\begin{equation}\nonumber
\begin{split}
& r_{x_0}=(R+r\cos\theta)\cos\phi, \quad r_x^{\prime}=\frac{\varepsilon}{2}(1-\cos n\phi)\cos\theta\cos\phi,\\
& r_{y_0}=(R+r\cos\theta)\sin\phi, \quad r_y^{\prime}=\frac{\varepsilon}{2}(1-\cos n\phi)\cos\theta\sin\phi,\\
& r_{z_0}=r\sin\theta, \quad r_z^{\prime}=\frac{\varepsilon}{2}(1-\cos n\phi)\sin\theta,
\end{split}
\end{equation}
wherein $R$ and $\phi$ are the large radius and azimuthal angle, $r$ and $\theta$ are the small radius and polar angle of a torus (see Figure~\ref{Surf}a), respectively, $\varepsilon$ is the amplitude of corrugation and $n$ denotes the period number of corrugation. The~position vector ${r}_0=(r_{x_0}, r_{y_0}, r_{z_0})$ describes a torus, ${r}=(r_x, r_y, r_z)$ describes the torus with corrugation shown in Figure~\ref{Surf}a,b, respectively. In addition, the vector ${r}^{\prime}=(r_x^{\prime}, r_y^{\prime}, r_z^{\prime})$ describes corrugation.

From the corrugated torus Equation~\eqref{Surf}, the two tangent unit basis vectors ${e}_{\eta}$, ${e}_{\xi}$ and the normal unit basis vector ${e}_{n}$ are obtained as
\begin{equation}
\begin{split}
& {e}_{\eta}=(-\cos\phi\sin\theta, -\sin\theta\sin\phi, \cos\theta),\\
&{e}_{\xi}=\frac{1}{U}(-T\sin\phi+S\cos\theta\cos\phi,\\
& \quad\quad\quad\quad\quad\quad T\cos\phi+S\cos\theta\sin\phi, S\sin\theta),\\
& {e}_{n}=-\frac{1}{U}(T\cos\theta\cos\phi+S\sin\phi,\\
& \quad\quad\quad\quad\quad\quad T\cos\theta\sin\phi-S\cos\phi, T\sin\theta),
\end{split}
\end{equation}
where
\begin{equation}\nonumber
\begin{split}
& Q=2r+\varepsilon-\varepsilon\cos n\phi,\\
& S=\varepsilon n\sin n\phi,\\
& T=2R+Q\cos\theta,\\
& U=\sqrt{S^2+T^2}.
\end{split}
\end{equation}

In addition, the position vector ${R}$ of a point close to the deformed torus is then described by
\begin{equation}
{R}={r}+q_3{e}_n,
\end{equation}
where $q_3$ is {the coordinate that measures the normal distance away from the surface}. According to the two position vectors ${r}$ and ${R}$, with the definitions of two matric tensors $g_{ab}=\partial_a{r}\cdot\partial_b{r}$ $(a, b=1, 2)$ and $G_{ij}=\partial_i{R}\cdot\partial_j{R}$ $(i, j=1, 2, 3)$, we obtain their covariant elements as
\begin{equation}\label{surfmetric}
g_{11}=1,\quad g_{22}=1, \quad
g_{12}=g_{21}=0,
\end{equation}
and
\begin{equation}\label{spacemetric}
\begin{split}
& G_{11}=(1-\frac{2T}{QU}q_3)^2+\frac{4S^2\sin^2\theta}{U^4}q_3^2,\\
& G_{12}=G_{21}=\frac{4S\sin\theta}{U^2}q_3-[\frac{T}{QU^3}\\
&\quad +\frac{(S^2+U^2)\cos\theta -\varepsilon n^2 T\cos n\phi}{U^5}]4S\sin\theta q_3^2,\\
& G_{22}=[1-\frac{(S^2+U^2)\cos\theta-\varepsilon n^2T\cos n\phi}{U^3}2q_3]^2\\
& \quad\quad\quad +\frac{4S^2\sin^2\theta}{U^4}q_3^2,\\
& G_{33}=1, \quad G_{13}=G_{31}=G_{23}=G_{32}=0,
\end{split}
\end{equation}
respectively. It is straightforward to check the relationship between $g$ and $G$ as
\begin{equation}\label{gGrelation}
G=f^2g,
\end{equation}
where $g$ and $G$ are the determinants of $g_{ab}$ and $G_{ij}$, respectively, and $f$ is the rescaling factor, that is
\begin{equation}\label{factor}
\begin{split}
& f=1-\frac{Q(S^2+U^2)\cos\theta-n^2\varepsilon QT\cos n\phi+U^2T}{QU^3}2q_3\\
& +\frac{(S^2+U^2)T\cos\theta -n^2\varepsilon T^2\cos n\phi-QS^2\sin^2\theta}{QU^4}4q_3^2.
\end{split}
\end{equation}

For a particle confined to the corrugated torus (see Figure~\ref{Surf}b), we deduce the effective Schr\"{o}dinger equation by the confining potential approach~\cite{Maraner1995A, Jaffe2003Quantum, Wang2018Geometric} (i.e., the thin-layer quantization scheme~\cite{Jensen1971, Costa1981Quantum, Wang2017Geometric}), that is
\begin{equation}\label{ESE}
-\frac{\hbar^2}{2m^*}(\frac{\partial^2}{\partial\eta^2} +\frac{\partial^2}{\partial\xi^2})\psi+V_g\psi=E\psi,
\end{equation}
where $\hbar$ is the Plank constant divided by $2\pi$, $m^*$ is the effective mass of electron, $\eta$ and $\xi$ are two tangent coordinate variables of the deformed torus, $\psi$ is the wave function describing the motion of the confined particle, $E$ is the energy with respect to $\psi$, and $V_g$ is the geometric potential induced by curvature, that is,
\begin{equation}\label{GP0}
\begin{split}
& V_g=-\frac{\hbar^2}{2m}\{\frac{4S^2\sin^2\theta}{U^4}\\
& +[\frac{Q(S^2+U^2)\cos\theta-T(U^2+n^2\varepsilon Q\cos n\phi)}{QU^3}]^2\}.
\end{split}
\end{equation}

\begin{figure}[htbp]
\centering
\includegraphics[width=0.18\textwidth]{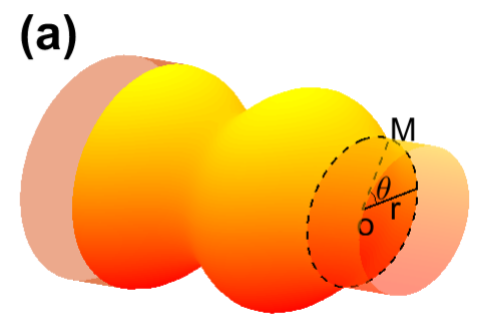}\quad
\includegraphics[width=0.18\textwidth]{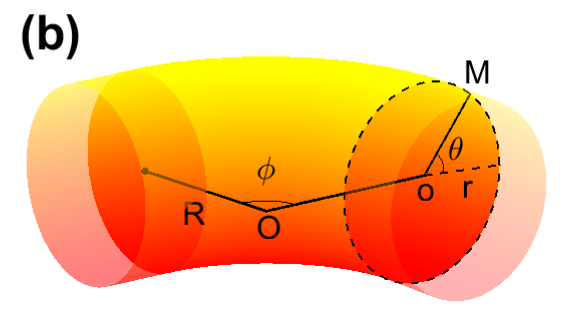}\\
\includegraphics[width=0.18\textwidth]{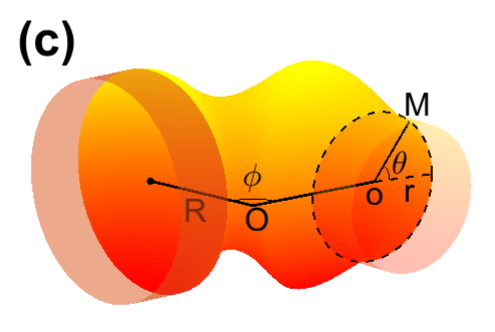}\quad
\includegraphics[width=0.18\textwidth]{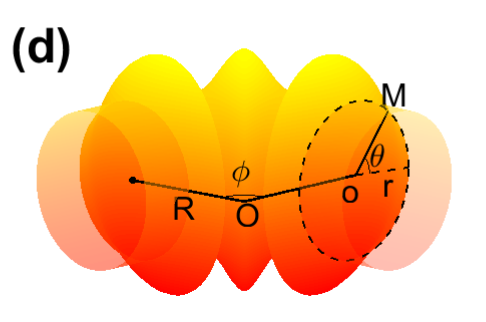}
\caption{(\textbf{a}) schematic of a deformed tube; (\textbf{b}) schematic of a quarter torus; (\textbf{c}) schematic of a quarter corrugated torus with $n=6$; (\textbf{d}) schematic of a quarter corrugated torus with $n=12$.}\label{Surf3}
\end{figure}

The geometric potential Equation~\eqref{GP0} is still valid to a part deformed torus with $\phi$ ranging from $\phi_0$ to $\phi_1$ ($0<\phi_0<\phi_1<2\pi$). In practical terms, we consider two models, one is a quarter corrugated torus with $n=6$ and the other with $n=12$ (see Figure~\ref{Surf3}c,d). For the sake of expressing convenience, a~deformed tube and a quarter torus are also sketched (Figure~\ref{Surf3}a,b). The~deformation of tube generates the geometric potential, like Dirac comb, attractive wells with identical depthes. The~Dirac comb can be used to filter particles with particular incident energies~\cite{Moraes2016Geometric}. It is obvious that the quarter torus can connect two tubes with identical radiuses in different directions. The~rest two quarter deformed toruses own the two features simultaneously. In what follows, we will discuss the properties of geometric potential of the quarter corrugated torus.

\begin{figure}[htbp]
\centering
\includegraphics[width=0.44\textwidth]{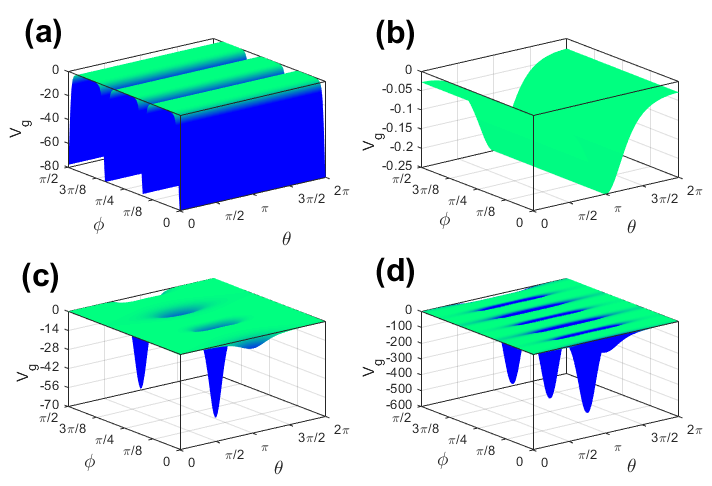}
\caption{(\textbf{a}) the geometric potential versus $\theta$ and $\phi$ for a corrugated tube of $r=2$ and $\varepsilon=1$; (\textbf{b}) the geometric potential versus $\theta$ and $\phi$ for a quarter torus of $R=4$ and $r=2$; (\textbf{c}) the geometric potential versus $\theta$ and $\phi$ for a quarter corrugated torus with $R=4$, $r=2$, $\varepsilon=1$ and $n=6$; (\textbf{d}) the geometric potential versus $\theta$ and $\phi$ for a quarter corrugated torus with $R=4$, $r=2$, $\varepsilon=1$ and $n=12$. Here, $\frac{\hbar^2}{2m^*}$ is taken as an unit.}\label{GPF}
\end{figure}

For a corrugated tube (see Figure~\ref{Surf3}a), the geometric potential is determined by the radius of tube, the periodic length and the amplitude of corrugation together. As the periodic length is fixed, it is easy to check that the amplitude is larger, and the attractive wells are deeper. When the amplitude of corrugation is fixed, it is easy to confirm that the periodic length is longer, and the wells are shallower. The~corrugation can create a Dirac comb for the particle passing the deformed torus. In order to learn the differences in the four cases described in Figure~\ref{Surf3}a--d, the corresponding four geometric potentials are sketched in Figure~\ref{GPF}a--d, respectively. In the case of a quarter torus, the geometric potential well is minimal at $\theta=\pi$, and it does not change by varying the azimuthal angle $\phi$. This result means that the inner side of the quarter torus is more capable of attracting particles. According to the figures of the geometric potential versus $\theta$ and $\phi$ shown in Figure~\ref{GPF}c,d, it is easy to {see} that the geometric potentials take their minima at $\theta=\pi$, $\phi=(i+1/2)\frac{\pi}{3}$ and $\phi=(i+1/2)\frac{\pi}{6}$ with $(i=0, 1, 2, \cdots)$ for $n=6, 12$, respectively. The~minima of the geometric potential can be in general expressed by
\vspace{12pt}
\begin{equation}\label{GPMinima}
V_{g_{m}}=-\frac{\hbar^2}{8m^*}[\frac{1}{R-r-\varepsilon}+\frac{1}{r+\varepsilon} -\frac{n^2\varepsilon}{2(R-r-\varepsilon)^2}]^2.
\end{equation}

It is obvious that the minima depends on curvature. For the quarter torus with corrugation, the geometric potential looks like the Dirac comb~\cite{Sakurai1993Modern} that can affect the transport of the electron confined to the particular quarter torus. When the variable $\theta$ is far away from $\pi$, the wells sharply become shallow. These results are useful to control the electronic transport by tailoring the quarter corrugated torus further. In what follows, we will consider the geometric effects of the quarter corrugated torus on transmission probability to illustrate the deduction.

\section{Transmission Probability in Periodically Corrugated Toruses}\label{section3}
\begin{figure}[htbp]
\centering
\includegraphics[width=0.36\textwidth]{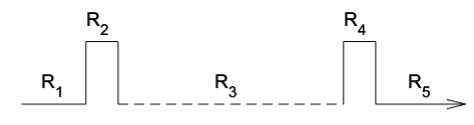}
\caption{A simplified model is sketched for a quarter corrugated torus connected with two skirts. $R_1$~describes the head skirt, $R_2$ and $R_4$ are two barriers, $R_3$ denotes a quarter torus with corrugations and $R_5$ stands for the foot skirt.}\label{Model1}
\end{figure}

In terms of the quarter corrugated torus, from left to right, the head skirt can be taken as a free electron beam source, the first connection between the head skirt and the deformed torus plays the role of a barrier, the central part is a quarter corrugated torus that can provides a geometric potential, the second connection between the central part and the foot skirt can be taken as the other barrier, and the foot skirt plays the role of a drain. This system in the $\phi$ direction can be simplified as a model that has the form as~\cite{Encinosa2000Surface, Wang2016Transmission} (see Figure~\ref{Model1}) which consists of two leads, two barriers and a geometric potential.

{By setting} the geometric potential {equal to zero}, the present model becomes {one} with double barriers. The~double barriers resonant tunneling structure is a typical microstructure that attracted various investigations~\cite{Li1990Noise, Chen1991Theoretical, Chen1991Dynamic, Kane1992Transmission, Doering1992Resonant, Nguyen2013Realization, Wang2017Quantum}. {A} quarter corrugated torus is {a mesostructure that has attracted much investigation, in which} the geometric potential induced by curvature will substantially affect the resonant peaks created by the double barriers. For transport properties, the resonant peaks correspond to quasibound states, which are associated with geometric quantum wells. The~electrons are primarily confined to a curved surface, and tend to stay in the attractive wells, but the electron has a certain probability to tunnel the corrugation, if the energy of the electron is nearly the energy eigenvale of the quansibound state. Physically, when the spatial dimension is reduced to a scale being comparable with the de Broglie wavelength of electron in the vicinity of Fermi energy, the wave nature of electron is expected to play an important role in transport properties. We will investigate four generic cases: a~periodically corrugated nanotube with $1.5$ periods, a quarter of a nanotorus, a quarter nanotorus with $1.5$ corrugations and a quarter nanotorus with three corrugations.

For the effective quantum equation Equation~\eqref{ESE}, when $\theta$ is fixed with constant, the equation can be simplified as
\begin{equation}\label{4SE1}
-\frac{\hbar^2}{2m^*}\frac{d^2}{d\xi^2}\psi(\xi) +V(\xi)\psi(\xi)=E\psi(\xi),
\end{equation}
where $m^*$ is the effective mass of electron, $\hbar$ is the Planck constant divided by $2\pi$, $\psi$ is a wave function describing the motion of electron confined to the surface with $\theta$ being a constant, $E$ is the energy with respect to $\psi$, and the potential $V(\xi)$ has the general form as
\begin{equation}\label{4Potential1}
V(\xi)=
\begin{cases}
& 0~\text{meV}, \quad \xi\in R_1,\\
& 20~\text{meV}, \quad \xi\in R_2,\\
& V_g(\xi), \quad \xi\in R_3,\\
& 20~\text{meV}, \quad \xi\in R_4,\\
& 0~\text{meV}, \quad \xi\in R_5,
\end{cases}
\end{equation}
wherein $V_g(\xi)$ denotes the geometric potential that is specifically expressed in Equation~\eqref{GP0} and meV stands for milli electron volts. This simplified model has the same form of the effective Hamiltonian of a corrugated surface~\cite{Wang2016Transmission}.%please confirm whether it is the unit, if is, please change all of them into not italic. Ps. we have changed it into not italic in the Eq.(13), please confirm.

Repeating the calculation process~\cite{Wang2016Transmission}, we first split $R_3$ up into segments instead of continuous variations of $V(\xi)$, in each segment $V_g(\xi)$ can be taken as a constant. Then, let us assume that $R_3$ consists of a sequence $N_3$ small segments, $R_2$ one segment ($N_2=1$) and $R_4$ one segment ($N_4=1$). It is straightforward to obtain that the total number of segments is $N=N_2+N_3+N_4$, and that of boundaries is $N+1$. It is worthwhile to notice that the tangent variable $\xi$ in the $j$-th segment of $R_3$ should be replaced by $\xi_j$, which can be described by
\begin{equation}\label{Xi}
\xi_j=\frac{1}{2}\int_{\phi_{j-1}}^{\phi_j}\sqrt{T^2 +S^2}d\phi,
\end{equation}
with $T=2R+(2r+\varepsilon-\varepsilon\cos n\phi)\cos\theta$ and $S=\varepsilon n\sin n\phi$. This integrated function is more complex than that of a corrugated surface~\cite{Wang2016Transmission}. {Based on the continuous conditions of the wave function and its first derivative}, the transmission probability can be calculated for electron moving on the particular~torus.

\begin{figure}[htbp]
\centering
\includegraphics[width=0.44\textwidth]{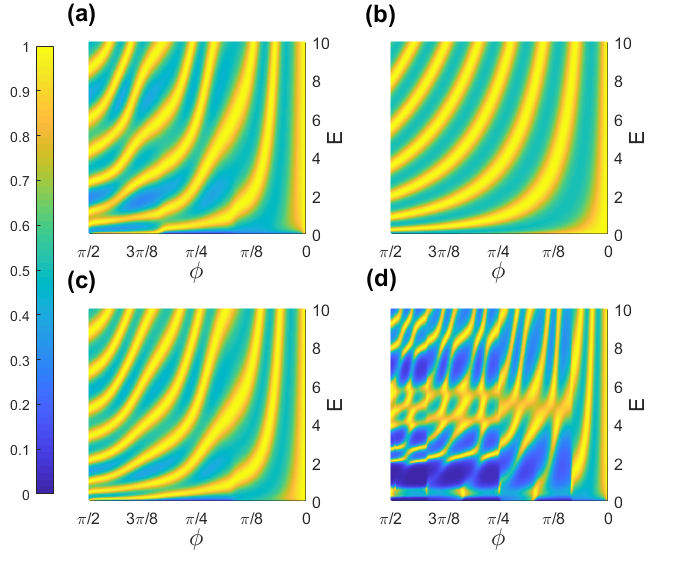}
\caption{Transmission probability versus incident energy $E$ and azimuthal angle $\phi$ for the model with $V(\xi)=20$ meV in $R_2$ and $R_4$, and for (\textbf{a}) $V_g(\xi)$ given by a corrugated tube; (\textbf{b}) $V_g(\xi)$ induced by a quarter torus; (\textbf{c}) $V_g(\xi)$ provided by a quarter torus with $1.5$ corrugations; (\textbf{d}) $V_g(\xi)$ provided by a quarter torus with three corrugations, and $m^*=0.173m_0$ and $\theta=\pi$ in $R_3$, $m^*=m_0$, otherwise.}\label{Fig5}
\end{figure}

\begin{figure}[htbp]
\centering
\includegraphics[width=0.44\textwidth]{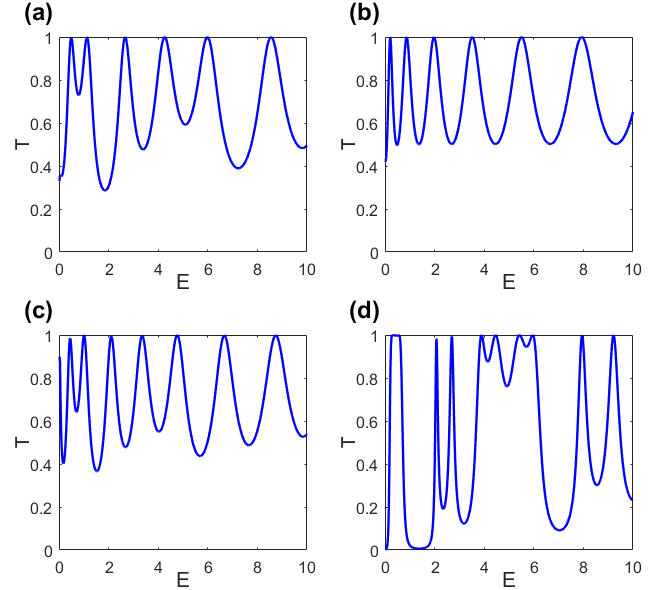}
\caption{Transmission probability versus incident energy $E$ for the model with \mbox{$V(\xi)=20$ meV} in $R_2$ and $R_4$, and for (\textbf{a}) $V_g(\xi)$ given by a corrugated tube; (\textbf{b}) $V_g(\xi)$ induced by a quarter torus; (\textbf{c}) $V_g(\xi)$ provided by a quarter torus with $1.5$ corrugations; (\textbf{d}) $V_g(\xi)$ provided by a quarter torus with $3$ corrugations, and $\theta=\pi$ and $m^*=0.173m_0$ in $R_3$, $m^*=m_0$, otherwise.}\label{Fig6}
\end{figure}

The transmission probabilities of the electron confined to the four curved surfaces described in Figure~\ref{Surf3}a--d are shown in Figure~\ref{Fig5}a--d, respectively. In these cases, the substrate is graphene in which the effective mass of electron is $m^*=0.173m_0$ with $m_0$ being the static mass of an electron. Figure~\ref{Fig5} shows that the geometric potential makes the strips of resonant tunneling peaks curved tend to the lower energy. In addition, the corrugation of nanotube and nanotorus deepens the geometric potential wells extremely, and the energy levels of the quasibound states shift downward, which means a shift of the resonant peaks to lower energy. As a result, there are more resonant peak strips shown in Figure~\ref{Fig5}.  Comparing Figure~\ref{Fig5}c with Figure~\ref{Fig5}a, we find that the bent corrugated nanotube can create the resonant peak strips more than the straight one. In addition, it is apparent that the resonant peak strips are deformed by corrugation. The~result is more significant than that of a corrugated surface~\cite{Wang2016Transmission}. The~deformation becomes more important when the number of corrugations is increased. As a consequence, the quarter corrugated nanotorus can be not only employed to connect different nanotubes, but also to control the electronic transport. These results become more straightforward when the two-dimensional (2D) figures describing the transmission probabilities versus the incident energy given by Figure~\ref{Fig6}. The~transmission gaps~\cite{Wang2016Transmission, Khelif2003Transmission, Chen2009Transmission, Chen2009Design} are deepened and emerged and the tunneling peaks~\cite{Liu1993Resonance, Guo1998Resonance, Zeng1999Resonant} are emerged and broadened (see Figure~\ref{Fig6}d).

\unskip

\begin{figure}[htbp]
\centering
\includegraphics[width=0.44\textwidth]{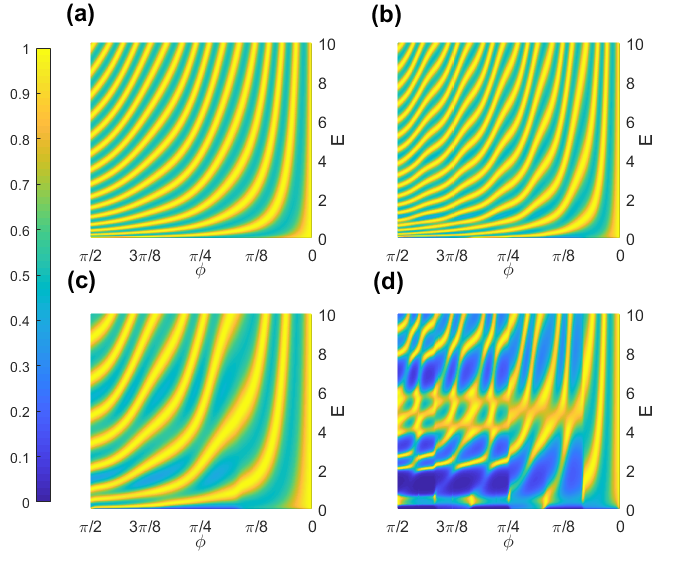}
\caption{Surface plot of the transmission probability as a function of $E$ and $\phi$ for $V(\xi)=20$ meV in $R_2$ and $R_4$ regions and in $R_3$ (a) $n=6$ with $\theta=\pi/2$; (\textbf{b}) $n=12$ with $\theta=\pi/2$; \mbox{(\textbf{c}) $n=6$} with $\theta=\pi$ and (\textbf{d}) $n=12$ with $\theta=\pi$, and $m^*=0.173m_0$, $m^*=m_0$, otherwise.}\label{Fig7}
\end{figure}

In order to learn the difference between the transmission probabilities for $\theta=\pi/2$ and $\theta=\pi$, the transmission probabilities are described in the plane constructed by $E$ and $\phi$ (see Figure~\ref{Fig7}). In the case of $\theta=\pi/2$, there are more resonant peak strips than that of the case of $\theta=\pi$. Those resonant peaks are generated by the boundaries constructed by different materials in which electrons have different effective masses. Comparing (c) and (d) with (a) and (b) in Figure~\ref{Fig7}, we find that the deeper wells are more capable of merging multiple resonant peaks into one. In other words, the geometric potential can broaden the width of the resonant peaks and the transmission gaps. The~result can be used to tailor the deformed torus to filter the electrons with special incident energies.

\begin{figure}[htbp]
\centering
\includegraphics[width=0.44\textwidth]{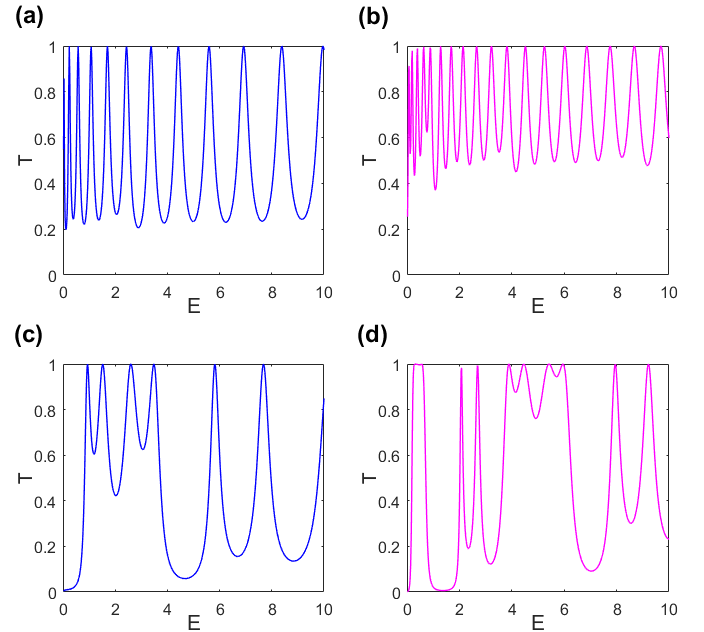}
\caption{Surface plot of the transmission probability as a function of $E$ with \mbox{(\textbf{a}) $m^*=0.067m_0$} and $\theta=\pi/2$; (\textbf{b}) $m^*=0.173m_0$ and $\theta=\pi/2$; (\textbf{c}) $m^*=0.067m_0$ and $\theta=\pi$; \mbox{(\textbf{d}) $m^*=0.173m_0$} and $\theta=\pi$ and $n=12$ in $R_3$, $m^*=m_0$ elsewhere, $V(\xi)=V_g(\xi)$ in $R_3$, $V(\xi)=20$ meV in $R_2$ and $R_4$ and $V(\xi)=0$ meV, otherwise.}\label{Fig8}
\end{figure}

For GaAs substrate with $m^*=0.067m_0$ and graphene $m^*=0.173m_0$, the transmission probability depending on the incident energy $E$ is described in Figure \ref{Fig8}, (a) $m^*=0.067m_0$ and $\theta=\pi/2$, (b)~$m^*=0.173m_0$ and $\theta=\pi/2$, (c) $m^*=0.067m_0$ and $\theta=\pi$ and (d) $m^*=0.173m_0$ and $\theta=\pi$. It is obvious that the above-mentioned results are more manifest in graphene than in the GaAs substrate. Generally, the geometric potential can influence the electronic transport of various materials. Specifically, for different materials, the geometric effects have a certain difference.
\unskip

\section{Conclusions}
In the present paper, the main result {has been} to obtain the effective quantum Hamiltonian for a particle confined to the corrugated torus. To illustrate the effects of the geometric potential on the electronic transmission, we numerically calculate the transmission probability. By researching four examples, a periodically corrugated nanotube, a quarter torus, a quarter torus with $1.5$ corrugations and a quarter torus with three corrugations, we found that the corrugations can extremely deepen the geometric potential wells. It is worthwhile to point out that the integer values of the number of corrugations for an usual torus would be broken when the investigated models are taken as a quarter torus. Subsequently, a part of corrugated torus can be selected to include any number of corrugations, and it can be used to connect two nanotubes with different radiuses in different directions.

The corrugations of the deformed nanotube and nanotorus can deepen the attractive wells, and the resonant peaks and valleys can emerge to broaden them.
%Please confirm this change
In the resonant peaks of the transmission probability, electrons {with particular incident energy} can easily pass, while, in the transmission gaps, electrons with specific incident energy are almost reflected. Practically, we can adjust the widths of the resonant peaks and the transmission gaps by designing corrugation, which may be used to design some quantum electronic and photonic nanodevices. In the present models, the interactions between electrons are neglected. A model including interactions of particles is still an interesting project that needs further investigation.

\section*{Acknowledgments}
This work is jointly supported by the National Major state Basic Research and Development of China (Grant No. 2016YFE0129300), the National Nature Science Foundation of China (Grants No. 11690030, No. 11475085, No. 11535005, No. 61425018). Y.-L. W. was funded by the Natural Science Foundation of Shandong Province of China (Grant No. ZR2017MA010).

\bibliographystyle{apsrev4-1}
\bibliography{myreferences}

\end{document}